\documentclass[10pt,twocolumn,twoside,letterpaper]{article}

\usepackage{dcolumn}
\usepackage{bm}
\usepackage[left=0.5in,right=0.5in,top=0.5in,bottom=1in]{geometry}%
\usepackage[utf8]{inputenc}
\usepackage[T1]{fontenc}
\usepackage{graphics}%
\usepackage{graphicx}
\usepackage{amsbsy}%
\usepackage{authblk}%
\usepackage{mathtools,stmaryrd,soul}%
\usepackage{amsmath,amssymb}
\usepackage{cuted}%
\usepackage{caption}%
\usepackage{xcolor}
\usepackage{mathptmx}

\begin{document}

\title{Reference-less complex wavefields characterization with a high-resolution wavefront sensor}

\author[1,2,3]{\bf Tengfei Wu}
\author[2,3]{\bf Pascal Berto}
\author[1,3,4,*]{\bf Marc Guillon}

\affil[1]{Université de Paris, SPPIN -- Saints-Pères Paris Institute for Neurosciences, CNRS, 75006 Paris, France}
\affil[2]{Sorbonne Université, CNRS, INSERM, Institut de la Vision, 17 Rue Moreau, 75012 Paris, France}
\affil[3]{Université de Paris, 75006 Paris, France}
\affil[4]{Institut Universitaire de France (IUF), Paris, France}

\affil[*]{marc.guillon@u-paris.fr}

\maketitle




\begin{strip}
\textbf{
Wavefront sensing is a widely-used non-interferometric, single-shot, and quantitative technique providing the spatial-phase of a beam. The phase is obtained by integrating the measured wavefront gradient.
Complex and random wavefields intrinsically contain a high density of singular phase structures (optical vortices) associated with non-conservative gradients making this integration step especially delicate.
Here, using a high-resolution wavefront sensor, we demonstrate experimentally a systematic approach for achieving the complete and quantitative reconstruction of complex wavefronts. 
Based on the Stokes' theorem, we propose an image segmentation algorithm to provide an accurate determination of the charge and location of optical vortices. This technique is expected to benefit to several fields requiring complex media characterization.}
\end{strip}

Quantitative phase mesurements of complex wavefields is at the basis of many applications including bio-imaging~\cite{naji,Betzig_NM_14,Jesacher_BioArxiv_21}, fibroscopy~\cite{Choi_PRL_12,Moser_OE_15}, diffractive tomography~\cite{Lauer_MST_08},  astronomy~\cite{Tyler_JOSA_2000}, 
and more generally for imaging through scattering media by phase conjugation~\cite{Gigan_NC_10,Yang_NP_15}.
When a coherent reference beam is available, complex wavefields may typically be measured using digital holography~\cite{Cuche1999,Matoba_DH_review_18}. 
Alternatively, several techniques have been developed for cases when no coherent reference beam is available, such as common-path interferometers~\cite{Marte_OE_06,Yang_APL_12,Marte_OE_14,So_Optica_17,Rosen_AOP_19}, point diffraction interferometry~\cite{Paterson_OC_08}, Shack-Hartman wavefront sensors (WFS)~\cite{Shack2001}, curvature sensing~\cite{Roddier_AO_88}, light-field imaging~\cite{Prevedel_NM_14}, phase diversity approaches~\cite{Khare_AO_15}, or phase retrieval algorithms~\cite{Fienup_AO_82,Paganin_PRE_01,Almoro2006}. All these techniques have their own advantages and try to optimize the compromise between ease of use and measurement reliability.
Among the latter techniques, WFS generally consist in a so-called ``Hartmann mask'' simply placed before a camera and have the specific advantages to be a robust single-shot technique offering a great ease of use. They have thus been the most used systems to measure distorted wavefronts arising from random refractive index inhomogeneities due to the atmosphere in astronomy, or due to tissues in bio-imaging. The main limitation associated to Shack-Hartmann WFS has long been the limited resolution associated with the low maximum density of micro-lens arrays. 
The advent of high-resolution WFS, either based on a modified Hartmann mask~\cite{Bon2009} or on a thin diffuser~\cite{Berujon_PRL_12, Berto_OL_17, Heidrich_OE_17, Heidrich_SR_19}, has represented an important step towards making WFS a promising alternative for measuring complex spatial phase profiles~\cite{Vellekoop_PRL_17}. 
However, despite their promising potential to measure complex WFs, high-resolution WFS have mostly been used, so far, for measuring smooth WF distortions, such as optical aberrations~\cite{naji,Betzig_NM_14} (typically projected onto Zernike polynomials~\cite{Love_AO_97}) or optical-path-length profiles of thin and transparent biological samples~\cite{Bon2009,Berto_OL_17}.
In contrast, complex and scattered wavefields comprise a high spatial density of phase singularities, namely optical vortices~\cite{Berry_PRSLA_74}, whose separation distance close to nucleation/annihilation centers can be arbitrarily small~\cite{Freund_PRL_94,Dennis_JPA_08, Pascucci_PRL_16}. 
These singular spiral phase structures are associated with non-conservative gradient fields that are especially challenging for WFS which, unlike interferometric methods, do not provide a direct measurement of the phase, but only measure phase gradients ${\bf g}$ (\emph{i.e.} the transverse component of the local wavevector). 
The problem of phase-spirals integration has appeared since the early ages of adaptive optics in astronomy~\cite{Fried_AO_92}. In this context, it has been shown that neglecting ``branch-cuts'' significantly degrades adaptive-optics performances~\cite{Tyler_JOSA_2000}. 
The identification of spiral phase structures in phase gradient maps ${\bf g}$ relies on their Helmholtz decomposition (HD)~\cite{LeBigot_OL_99,Tikhomirova_JOSA_02}. According to Helmholtz's theorem~\cite{Bremer_IEEE_13}, the vector field ${\bf g}$ can be split into an irrotational (curl-free) component and a solenoidal (divergence-free or rotational) component~\cite{Fried_JOSA_98,LeBigot_OL_99,Tyler_JOSA_2000,Senthilkumaran_JOSAA_12}. 
Most of current integration techniques for WFS basically consist in computing $\nabla\cdot {\bf g}$, 
so implicitly canceling the solenoidal component of the vector field~\cite{Huang2015}. The vector potential associated with the solenoidal component (so-called ``branch-point'' potential~\cite{Fried_JOSA_98,LeBigot_OL_99}) exhibits peaks at vortex locations, so allowing their localization and reconstruction~\cite{Tyler_JOSA_2000}. However, although WFS have demonstrated their ability to detect and characterize optical vortices~\cite{Fried_AO_92,Dainty_OE_12,Fallah_AO_15}, complete wavefront reconstruction by vortex-gradients integration has given rise to many works considering only simple experimental cases involving a single vortex~\cite{Soldatenkov_OL_07,Dainty_OE_10,Bai_OE_15}, or more vortices but in numerical simulations~\cite{Fried_AO_92,Fried_JOSA_98,Tyler_JOSA_2000,Olivier_JOSA_07,Soldatenkov_OL_07}. 
We believe that the high density of singular phase structures in complex wavefields has probably been the main obstacle for measuring experimentally complex wavefronts with a WFS. 
In WF shaping experiments, neglecting a single optical vortex is equivalent to adding a complementary spiral phase mask, which has been described as yielding a two-dimensional Hilbert transform of the field~\cite{Larkin_JOSA_01}. For complex speckled wavefields, such a single spiral transform induces a major change in patterns since resulting in an inversion of intensity contrasts~\cite{Gateau_PRL_17, Gateau_Optica_19}.
Here, we propose a robust and systematic numerical data-processing technique and demonstrate the experimental quantitative reconstruction of complex wavefronts containing up to $133$ optical vortices with a high-resolution WFS based on a $1.3~{\rm MP}$ camera~\cite{Berto_OL_17}. The camera sensor is thus similar to those typically used by regular WFS based on micro-lens arrays. Wavefront reconstruction was achieved using a segmentation algorithm that optimizes the identification of nearby optical vortices. According to the Stokes' theorem~\cite{Fried_JOSA_98}, segmentation optimizes the computation areas to measure the charge of vortices, based on the integration of the ``branch-point'' potential.

\begin{figure}[t]
\centering
\includegraphics[width=\linewidth]{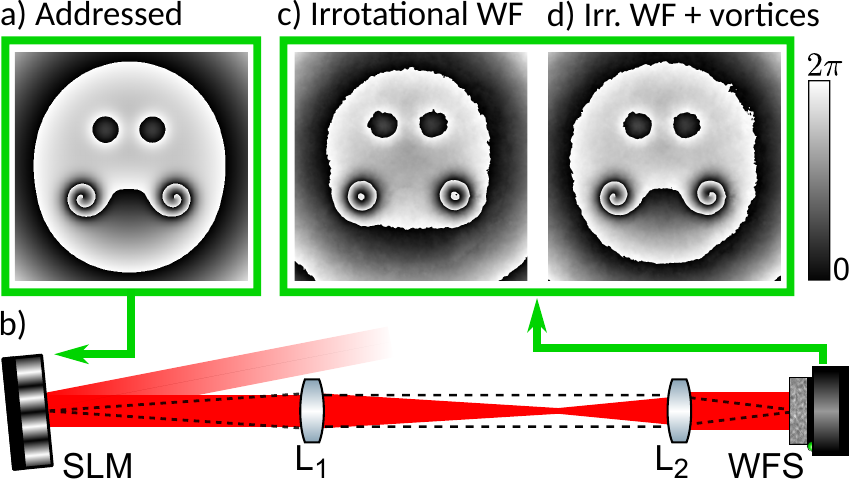}
\caption{ \label{fig:fig1} \textbf{Experimental vortex WF sensing.} A phase profile containing optical vortices~(a) is addressed onto a spatial light modulator (SLM)~(b) illuminated by a colimated laser beam and imaged onto a high-resolution wavefront sensor (WFS) with a telescope ($L_1$, $L_2$). Considering only the irrotational component of the gradient field detected by the WFS leads to an erroneous WF lacking optical vortices~(c), unlike full Helmholtz decomposition (HD) of the gradient field~(d).}
\end{figure}

To summarize, the problem to rebuild singular WFs is mostly threefold. 
First, the phase-gradient map ${\bf g}$ measured by a WFS is a non-conservative vector field defined over a multiply connected domain. Direct spatial integration is thus not possible since the integral value depends on the line-path. 
Second, optical vortices are associated with an infinite phase-gradient ${\bf g}$ at the singularity locations, which appears as critically incompatible with WFS. 
Third, WFS may imprecisely measure strongly fluctuating gradients around anisotropic vortices, leading to inaccurate vortex characterization, in particular their spiral spectrum~\cite{Torner_OE_05}. Such cases especially appear in random wavefields, wherein pairs of vortices of opposit charges are frequently close to one another~\cite{Freund_PRL_94}. This configuration results in large phase gradients in between vortices and makes the computation of the charges (the circulation of ${\bf g}$) especially delicate. The image segmentation we propose more specifically addresses this latter challenge~\cite{Patent_vortex_2020}.

By way of illustration, a phase pattern exhibiting optical vortices has been designed (Fig.~\ref{fig:fig1}a) and addressed to a phase-only spatial light modulator (SLM) (Hamamatsu, LCOS-X10468-01), illuminated by a spatially filtered, polarized, and collimated laser beam (Fig.~\ref{fig:fig1}b). The SLM allows displaying patterns exhibiting both smooth local WF distortions (such as lenses for the eyes and the face contour for instance) as well as optical vortices of any topological charge (left- and right-handed optical vortices at the tips of the curling mustache). The phase-gradient map is detected with a custom-built high-resolution WFS comprising $\simeq 2\times 10^4$ phase pixels and described in Ref.~\cite{Berto_OL_17}.
The WFS is conjugated to the SLM with a Galilean telescope in a $4-f$ configuration. 
The full Helmholtz decomposition (HD) of the gradient vector-field detected by the WFS can be achieved according to:
\begin{equation}
{\bf g}=\nabla \varphi_{ir}+\nabla \times {\bf A}
\label{eq:helmholtz}
\end{equation}
so splitting appart the irrotational contribution of the regular phase $\varphi_{ir}$ and the solenoidal contribution of a vector potential ${\bf A}$. The sought-for complete phase profile $\varphi$, whose gradient-field is ${\bf g}$, can then be written as $\varphi=\varphi_{ir}+\varphi_s$. 
Typical direct numerical integration~\cite{Huang2015} yields the regular WF shown in Fig.~\ref{fig:fig1}c, missing phase singularities because the non-conservative (solenoidal) contribution to the WF-gradient has been ignored. 
The singular phase contribution $\varphi_s$ (or ``hidden phase''~\cite{Fried_JOSA_98}), is defined over a multiply-connected domain, and satisfies $\nabla\varphi_s=\nabla\times{\bf A}$. Solving this latter equation then allows proper reconstruction of the WF (Fig.~\ref{fig:fig1}d). 

\begin{figure*}[t]
\centering
\includegraphics[width=0.8\textwidth]{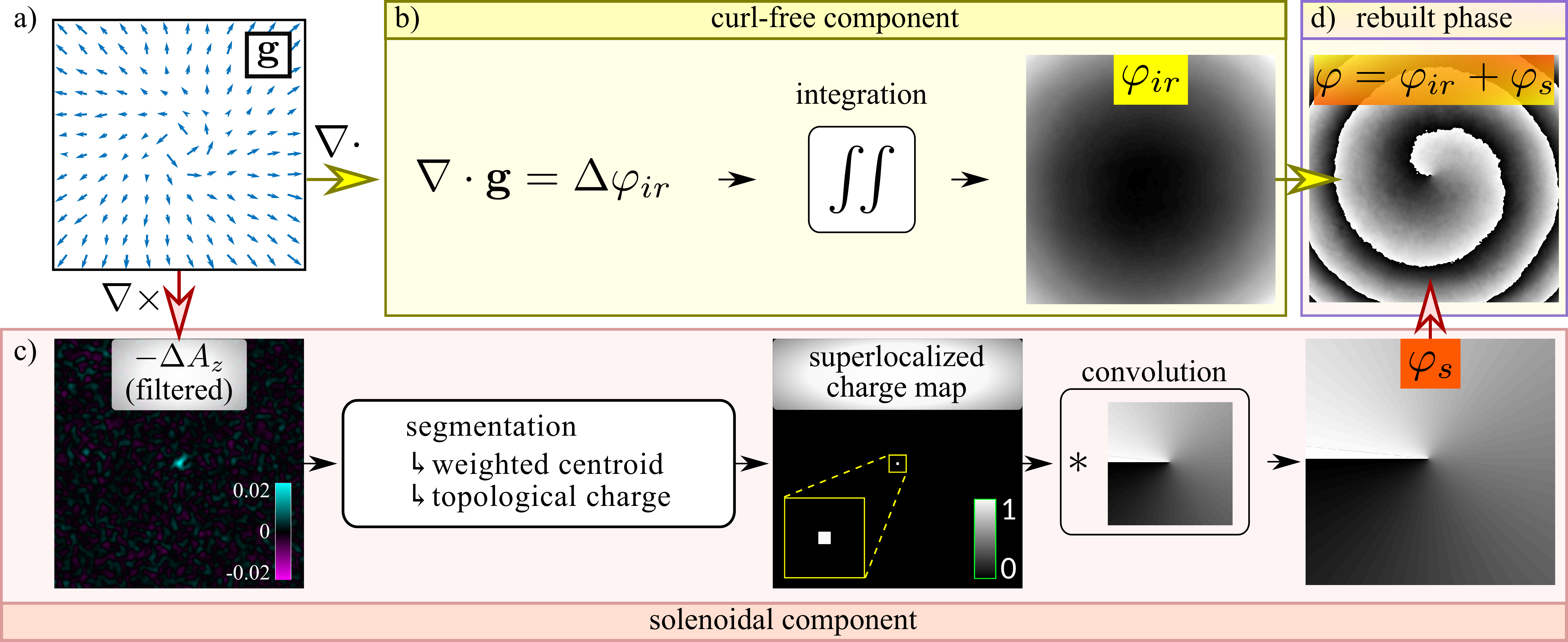}
\caption{ \textbf{Principle of full WF reconstruction.} The divergence and the curl of the phase gradient map ${\bf g}$ (a) are computed to extract the irrotational phase $\varphi_{ir}$ (b) and the solenoidal phase $\varphi_s$ (c). Double space integration of $\nabla\cdot{\bf g}$ yields the irrotational phase (a parabola in (b)). The curl of ${\bf g}$ yields $-\Delta {\bf A}$ where ${\bf A}$ is the potential vector of the Helmholtz decomposition. Image segmentation and weighted centroid computation of the peaks in $-\Delta {\bf A}$ then allows reconstructing a Dirac-like distribution whose convolution by a spiral phase profile yields $\varphi_s$. The complete phase $\varphi$ is finally rebuilt by summing the two components $\varphi_{ir}$ and $\varphi_s$ (d).
}
\label{fig:fig2}
\end{figure*}

We now detail the steps allowing us to achieve rigorous HD (See also Supp. Mat.). 
First, taking the curl of Eq.~\eqref{eq:helmholtz}, it appears that the potential vector ${\bf A}$ is a solution of the equation: $\nabla\times {\bf g}=\nabla (\nabla\cdot {\bf A})-\Delta {\bf A}$, where "$\Delta =\nabla^2$" denotes the Laplacian operator. Determining the potential vector thus requires fixing a gauge. The Coulomb gauge $\nabla \cdot {\bf A}=0$ is chosen here for obvious convenience~\cite{Fried_JOSA_98,LeBigot_OL_99,Tyler_JOSA_2000}. Since ${\bf g}$ is a two-dimensional vector field (in say $x$, $y$ plane), we may then write ${\bf A}=A_z{\bf e_z}$ without loss of generality. 
Second, introducing the circular vector $\pmb{\sigma_-}={\bf e_x}-i{\bf e_y}$, simple manipulation of Eq.~\eqref{eq:helmholtz} yields: ${\bf g} \cdot \pmb{\sigma_-} = g_x+ig_y= (\partial_x+i\partial_y)(\varphi_{ir}-iA_z)$. The HD can then be efficiently achieved numerically thanks to a single computation step:
\begin{equation}
\varphi_{ir}-iA_z={\cal F}^{-1}\left\{ \frac{{\cal F}\left[{\bf g}\cdot \pmb{\sigma_-}\right]}{i{\bf k}\cdot \pmb{\sigma_-}}\right\}
\label{eq:integration}
\end{equation}
where ${\bf k}$ stands for the two-dimentional coordinate vector in the reciprocal Fourier space.
The regular phase component $\varphi_{ir}$ is thus recovered the same way as previously proposed~\cite{Huang2015,Bon_AO_12} (Fig.~\ref{fig:fig2}b).
Third, for complete HD over a bounded domain, the contribution of a so-called additional ``harmonic'' (or translation) term ${\bf h}$ must be considered when the flow through the boundary is not zero~\cite{Bremer_IEEE_13}. This translation term, accounting for a global tip/tilt of the WF, is both curl-free and divergence-free. If computing derivation and integration through discrete Fourier transforms without care, implicit periodic boundary conditions cancels out this term. Here we thus retrieve the term ${\bf h}$ by symmetrizing the gradient field ${\bf g}$ prior to gradient integration (Eq.~\eqref{eq:integration}) as performed in Ref.~\cite{Bon_AO_12}, which conveniently includes ${\bf h}$ in the curl-free component ($\nabla\varphi_{ir}$) of the HD.
The divergence-free component requires futher processing steps to obtain the singular phase pattern $\varphi_s$ from the potential-vector component $A_z$, as detailed hereafter.

\begin{figure*}[ht]
\centering
\includegraphics[width=\textwidth]{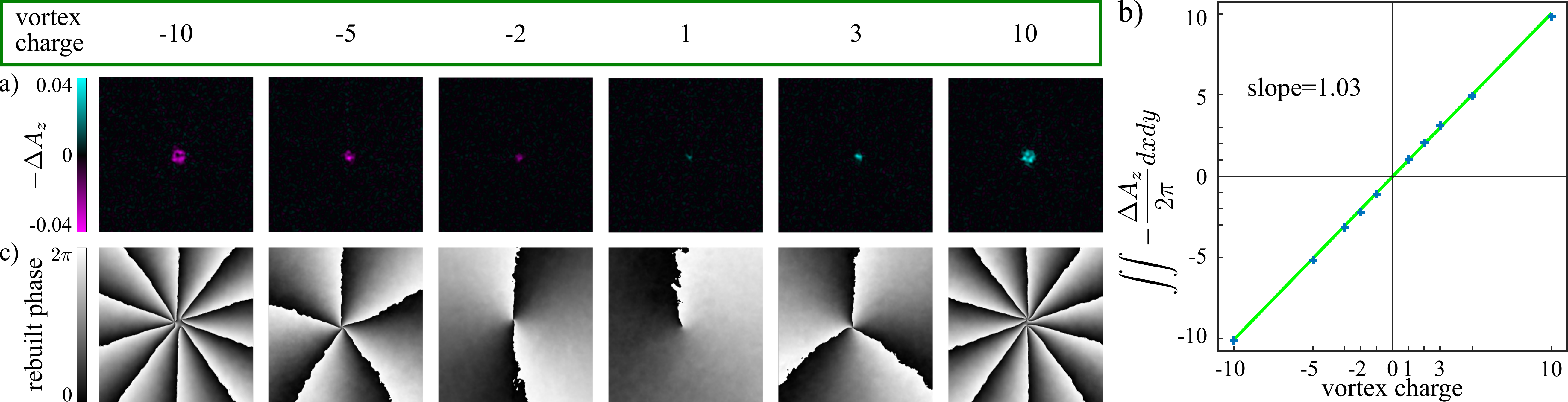}%
\caption{ \textbf{Reconstruction of phase spirals of various topological charges $n$ from $-10$ to $+10$.} The potential vector ${\bf A}$ of a spiral phase of topological charge $1$ is, in theory, the Green function of Laplace equation. In practice, $\Delta A_z$ exhibit peaks at the vortex location (a) whose size depend on the topological charge of the corresponding vortex (in (a), $\Delta A_z$ has been slightly filtered for the sake of readability). After segmentation, the integral computation of this peak yields the expected value $2\pi n$ (with a $3\%$ precision) (b). Phase profiles are then rebuilt~(c).%
}
\label{fig:fig3}
\end{figure*}

A vortex is essentially characterized by the circulation of the vector flow around the singularity, namely its topological charge (or winding number) defined from the circulation of ${\bf g}$ around the vortex. Applying Stokes' theorem to the definition of $n$ yields:
\begin{equation}
n = \frac{1}{2\pi}\displaystyle{\oint_{\cal C}} {\bf g}\cdot {\rm d}\pmb{ \ell} =- \frac{1}{2\pi }\displaystyle{\int_{\cal S}} \Delta A_z {\rm d}S
\label{eq:Stokes}
\end{equation}
Reducing the contour length ${\cal C}$ (and so the enclosed surface ${\cal S}$) to zero, it appears that $-A_z/(2\pi n)$ is the Green function of the two-dimensional Laplace equation. In theory, $-\Delta A_z/(2\pi n)$ is thus a Dirac distribution, making it easy to identify optical vortices~\cite{Fried_JOSA_98,LeBigot_OL_99,Fallah_AO_15} (see Fig.~\ref{fig:fig2}c). In principle, the corresponding sought-for singular phase component~$\varphi_s$ could then be simply obtained by convolving $-\Delta A_z/(2\pi n)$ with a single $+1$ optical vortex. However, in practice, rebuilding $\varphi_s$ this way is not possible. The main difficulty is that the experimental $-\Delta A_z/(2\pi n)$ map is not a perfect Dirac distribution (or a single-pixeled non-zero data map) for three main reasons: first, experimental data are affected by noise, second, ${\bf g}$ is filtered by the optical transfer function of the WFS and third, optical vortices are associated with a vanishing intensity that compromises accurate gradient measurement in their vicinity. As detailed in Ref.~\cite{Berto_OL_17}, the optical transfer function of a WFS is especially limited by the non-overlapping condition, which imposes a maximum magnitude for the eigenvalues of the Jacobi matrix of ${\bf g}$ (\emph{i.e.} the Hessian matrix of $\varphi$). As a first consequence, the large curvatures of the phase component $\varphi_r$ (\emph{i.e.} its second derivatives) may be underestimated. As a second consequence, the diverging magnitude of ${\bf g}$ (as $1/r$) prevents its proper estimation at close distances $r$ from the optical vortex location, as well as the estimation of the Hessian coefficients $\partial_x g_y$ and $\partial_y g_x$. Therefore, the measurement of the vector potential $A_z$ is wrong in the vicinity of the vortex center and the obtained peak is not single-pixeled (Fig.~\ref{fig:fig2}c).

\begin{figure}[b!]
\centering
\includegraphics[width=\linewidth]{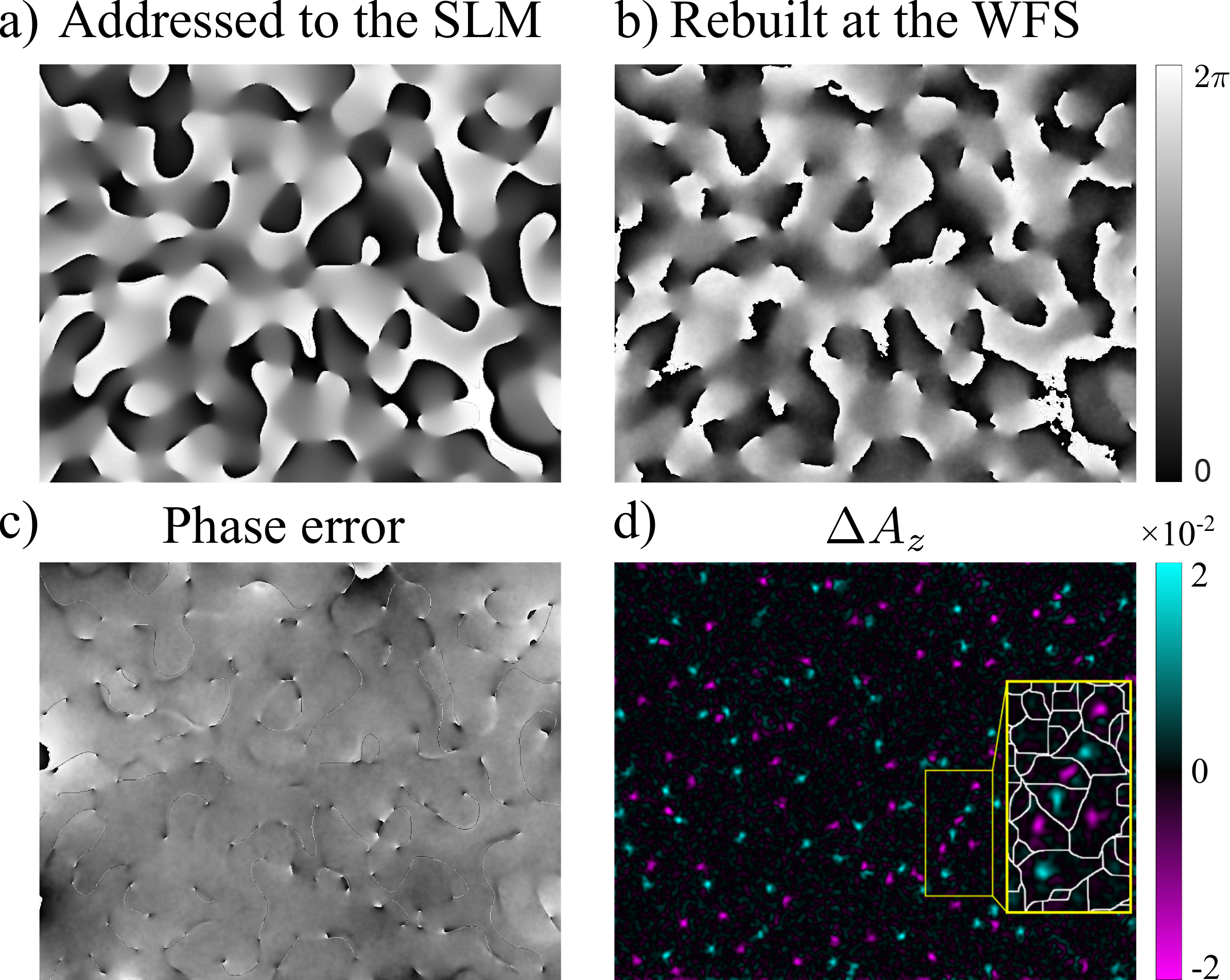}
\caption{ \textbf{WF reconstruction of a complex wavefield.} The phase pattern addressed to the SLM (a) is retrieved thanks to the WFS (b) after removal of the first few low-order aberrations (tip/tilt, defocus and astigmatism). The phase difference in (c) clearly demonstrates the efficient wavefront reconstructions, especially the non-conservative part. Remaining smooth distortions in $\varphi_{ir}$ at the boundaries are partly attributed to inaccurate phase modulation by the SLM. The (filtered) $\Delta {\bf A_z}$ map (d) allows identifying the $133$ vortices. The inset shows the optimized segmentation of nearby vortices.}
\label{fig:fig4}
\end{figure}

In experiments, the precise identification of the location and the charge of optical vortices from the computed $\Delta A_z$-map thus demands an image processing step. Notably, the circulation of ${\bf g}$ in Eq.~\eqref{eq:Stokes} can yield an accurate measure of the vortex charge provided that the contour ${\cal C}$ surrounds the vortex at a large enough distance, so that ${\bf g}$ is small enough to be accurately measured by the WFS. Consequently, although the estimate of $\Delta A_z$ is wrong in the vicinity of the vortex, the peak integral achieved over a large enough surface ${\cal S}$ (enclosed by ${\cal C}$) provides the proper charge, under the Stokes' theorem (Eq.~\eqref{eq:Stokes}). 
Vortices may be characterized by a simple peak detection algorithm, after regularization of the $\Delta A_z$-map with a low-pass filter~\cite{Tikhomirova_JOSA_02}. However, this solution is not robust since it requires setting manually an integration radius. Such an adjustment of the integration radius is all the more impractical when considering two typical cases: either high densities of optical vortices potentially located at arbitrarily small distances (see Fig.~\ref{fig:fig4}d), or high-order vortices, associated with charge-dependent peak-sizes in the $\Delta A_z$-maps (see Fig.~\ref{fig:fig3}a).
In contrast, the solution we propose consists in performing an image segmentation that automatically optimizes the integration surface by adjusting it to the size and density of peaks. The main processing steps are summarized in Fig.~\ref{fig:fig2}c (See also Supp. Mat.). 
The segmentation of the $\Delta A_z$-map is achieved in two substeps. 
First, $|\Delta A_z|$ is filtered in order to avoid oversegmentation, with a Gaussian filter $G$ whose waist is chosen equal to the WFS resolution~\cite{Berto_OL_17} ($\simeq 8$ camera pixels). 
Second, a watershed operation (matlab\textsuperscript{\tiny\textregistered}, \emph{Image Processing Toolbox}) is applied to $-G\ast|\Delta A_z|$. Noteworthy, we observed that segmentation of $-G\ast|\Delta A_z|$ rather than $-|G\ast\Delta A_z|$ more efficiently cancels out pairs of vortices too close to be resolved by the WFS. The resulting segments are thus structured according to extrema of $\Delta A_z$, the latters corresponding to positive and negative vortices locations.
Integration and weighted-centroid computation of the unfiltered $\Delta A_z$ image over each segment yields the charge and the precise location of each vortex. To obtain $\varphi_s$, a Dirac-like vortex map is rebuilt based on the result of the former step, and convolved by a $+1$ spiral phase mask (see Fig.~\ref{fig:fig2}c). Finally, the complete phase reconstruction $\varphi=\varphi_{ir}+\varphi_s$ is computed and wrapped between $0$ and $2\pi$ (Fig.~\ref{fig:fig2}d). 
Even if the proposed image processing algorithm includes a filtering operation, it is of a very different nature from previously suggested regularization approaches~\cite{Tikhomirova_JOSA_02}, because the filter size is set according to the WFS resolution (to avoid oversegmentation) and not to the vortices to be characterized.
Oversegmentation was observed to degrade charge-measurement reliability and lengthen the processing time, measured to be $0.54s$ on an Intel\textregistered~Core i5-9400H CPU for a $1.3~{\rm MP}$ map at maximal vortex density. %

To demonstrate the flexibility of this approach to characterize and rebuild optical vortices of any charge, we addressed phase spirals with charges ranging from $-10$ to $+10$ (Fig.~\ref{fig:fig3}). Because of the diverging phase gradient at the vortex center, the $\Delta A_z$ maps exhibit peaks whose widths increase with the charge $n$ of the vortex (Fig.~\ref{fig:fig3}a). Nevertheless, integration of $-\Delta A_z/(2\pi)$ over segmented regions yields $n$ within a $3\%$ accuracy range (Fig.~\ref{fig:fig3}b). Except for specially designed optical vortex beams carrying fractional charges~\cite{Vasnetsov_JOA_04}, for beams propagating in free space, the topological charge is typically an integer. Rounding the integral to the closest integer value then allows an accurate reconstruction of the optical vortices (Fig.~\ref{fig:fig3}c). 
Differences between the rebuilt phase profiles and the perfect ones addressed to the SLM are due to the contribution of $\varphi_{ir}$ arising from uniformity imperfections of the SLM.%

Finally, we demonstrate the possibility to retrieve the phase of complex random wavefields with our algorithm. Random wavefields contain a high density of optical vortices of charge $+1$ and $-1$~\cite{Longuet_Higgins_JOSA_60, Berry_PRSLA_74,Freund_1001_correlations}. 
These vortices exhibit elliptical phase and intensity profiles along the azimuthal coordinate. The non-uniform increase of the phase around the singular point may then alter the ability to detect them if the phase-gradient magnitude is locally too large. Furthermore, the separation distance between vortices may be much smaller than the speckle grain size, especially when close to creation or annihilation events of pairs of vortices~\cite{Freund_PRL_94,Nye_PRSLA_88,Pascucci_PRL_16}. 
Such a complex wavefield was numerically generated by taking the Fourier transform of a random phase map of finite aperture, and addressed to the SLM (Fig.~\ref{fig:fig4}a). Despite the aforementioned specific difficulties, the WF could be efficiently rebuilt (Fig.~\ref{fig:fig4}b). The accurate reconstruction can be visually appreciated by considering $0$-$2\pi$ equiphase lines, easy to identify with a gray-level colormap as abrupt white-to-black drops. The difference between the rebuilt and the input phase profiles is shown in Fig.~\ref{fig:fig4}c demonstrating the almost perfect reconstruction of the $133$ optical vortices. The six lowest order Zernike aberration modes were removed (piston, tip/tilt, defocus, astigmatisms) for better visualization.
Again, differences mostly appear on the $\varphi_{ir}$ contribution on the edges of the SLM, where the SLM reliability degrades and where aberrations introduced by the relay telescope are maximum. The dense experimental map of vortices distribution and charges is shown in Fig~\ref{fig:fig4}d.

Relying on a high-resolution WFS, we could thus propose a systematic and robust approach to rebuild complex wavefields containing a high density of optical vortices. The proposed method first consists in performing a HD of the local wavevector field ${\bf g}$ measured by the WFS. Importantly, the circulation of ${\bf g}/(2\pi)$ over vortices, computed as the integral $\nabla \times {\bf g}$ over large enough surface areas (under the Stokes' theorem), yields the topological charge of vortices. The systematic reconstruction of the optical vortex map relies on an image segmentation that optimizes surface integration and thus, vortex identification. The robustness of phase-spiral-reconstructions further relies on the quantization prior about the detected topological charges. Our method is applicable to any high-resolution WFS or system requiring integrating the local wavevector map.
Full reconstruction of WFs with a WFS represents an important step to make WFS efficient reference-less spatial-phase detectors and to allow random wavefields characterization, in particular with incoherent light sources. These developments are of interest for applications such as adaptive optics, diffractive tomography, as well as beam shaping behind scattering and complex media. 

\section*{Funding Information}
This work was partially funded by the french Agence Nationale pour la Recherche (SpeckleSTED ANR-18-CE42-0008-01), by the technology transfer office SATT/Erganeo (project 520 and project 600) and by Region île de France (DIM ELICIT, 3-DiPSI).
\section*{Acknowledgments}
The authors thank Jacques Boutet de Monvel and Pierre Bon for careful reading of the manuscript, and Benoit Forget for stimulating discussions.

\section*{References}



\providecommand{\noopsort}[1]{}\providecommand{\singleletter}[1]{#1}%

\end{document}